\documentclass[letterpaper,aps,preprintnumbers,longbibliography,notitlepage,longbibliography]{revtex4-1}
\usepackage[latin9]{inputenc}
\usepackage{amsmath}
\usepackage{amssymb}
\usepackage{graphicx}
\usepackage{esint}
\usepackage[scr=boondoxo, scrscaled=1.05]{mathalfa}
\allowdisplaybreaks

\makeatletter

\pdfpageheight\paperheight
\pdfpagewidth\paperwidth

\pdfoutput=1

\usepackage{epsfig}
\usepackage{graphics}

\usepackage{mciteplus}
\mciteErrorOnUnknownfalse

\makeatother

\begin{document}

\preprint{FERMILAB-PUB-18-088-A}

\title{Value of the Cosmological Constant in Emergent Quantum Gravity}

\author{Craig  Hogan}
\affiliation{University of Chicago and Fermilab}

\begin{abstract}
It is suggested that the exact value of the
cosmological constant could be derived from first principles, based on entanglement of  the Standard Model field vacuum with emergent  holographic quantum geometry. For the observed value of the cosmological constant, geometrical  information  is shown to agree closely with the spatial information density of the QCD vacuum, estimated in a free-field approximation.  The comparison is motivated  by a model of exotic  rotational fluctuations in the inertial frame that can be precisely tested in laboratory experiments.   Cosmic acceleration in this model is always positive, but fluctuates with characteristic coherence length $\approx 100$km and bandwidth $\approx 3000$ Hz.  
\end{abstract}
\maketitle

The cosmological constant $\Lambda$ was introduced  by Einstein just over a century ago into the fundamental equations of general relativity. 
Its  physical effect  is to accelerate  the expansion of empty space: in the absence of any form of gravitating matter, two test particles at separation $r$ accelerate apart at a rate
\begin{equation}\label{accelerate}
\sqrt{\ddot r / r }=\sqrt{ \Lambda/3} \equiv H_\Lambda.
\end{equation}
In classical theory,  it is an arbitrary parameter that can take any value. 
Its value in the real world is determined by cosmological measurements.


Acceleration of  cosmic expansion in the real universe was  actually found just over twenty years ago\cite{Riess:1998cb,Perlmutter:1998np}, and since then, a nonzero value of $\Lambda$ has been widely adopted as a standard ingredient in the simplest and most successful (so-called ``$\Lambda$CDM'') cosmological models.
Its  value has now been measured \cite{Riess:2016jrr,Olive2014} to a precision of about five percent: a typical fit to current cosmological data gives
\footnote{This estimate corresponds to a combination of standard parameters
$H_\Lambda=  \Omega_\Lambda^{1/2} H_0$, where $H_0$ denotes the current value of the Hubble constant and $\Omega_\Lambda$ denotes the density parameter of the cosmological constant.  The value quoted is derived by using  $h_0= 0.73\pm 0.017$ and $\Omega_\Lambda= 0.7\pm 0.01$ in standard cosmological notation\cite{Riess:2016jrr,Olive2014}.}
\begin{equation}\label{ccvalue}
\Lambda= 2.94\pm 0.15 \times 10^{-122} t_P^{-2},
\end{equation}
in natural units of the Planck time, $t_P\equiv \sqrt{\hbar G/c}= 5.4\times 10^{-44}\,$sec, set by  Planck's quantum constant $\hbar$,  Newton's gravitational constant $G$,  and the speed of light $c$.


It is  remarkable that with a few basic symmetries and  a small number of parameters including $\Lambda$, the  $\Lambda$CDM cosmological model precisely  accounts for many detailed features of cosmic structure and evolution\cite{Abbott:2017wau}.
However, there is no generally accepted theory that makes sense of   why $\Lambda$ is not  exactly zero, why it is so small in natural units, or why it has the particular value it does. 
In quantum field theory, the natural value of $\Lambda$  is of the order of unity, not $10^{-122}$. One widely held view is that there is no way to calculate its value precisely---  it is set at random, and its actual value in our universe is determined by  anthropic selection\cite{Weinberg:1988cp}.

There are of course other very large numbers in physics, and some of them have good explanations. For example, the  strong interactions of the Standard Model fields become strong at a  scale comparable  to the mass of the pion, $m_\pi\approx 10^{-20} m_P$, where $m_P\equiv  \sqrt{\hbar c/ G}= 2.1\times 10^{-8}\,$ kg denotes the Planck mass. The exponentially  large ratio arises in quantum  field theory  from the logarithmic running with energy of a renormalized coupling constant\cite{Wilczek1999}. 
It has long been suspected (e.g. \cite{Z68}) that there might similarly be a deep physical reason for a relationship of the form 
$\Lambda \approx  m_\pi^6$ in Planck units.

It is proposed here that a unified,  emergent  quantum theory of space, time, and fields  may provide a new, precise and testable way to explain how the value of $\Lambda$ relates to particle physics. 
The cosmological constant differs from zero because  an exact flat symmetry of  empty space is broken slightly by entanglement of exotic geometrical states with the  Standard Model field vacuum.

If quantum gravity behaved like particles and fields, it would not be important on scales much larger than the Planck length, $\l_P\equiv  \sqrt{\hbar G/ c^3}= 1.6\times 10^{-35}\,$m. However,  compelling thought experiments have  long  supported  the need for new, exotic geometrical correlations on all scales.  The entropy $S_{BH}= A/4 l_P^2$ of  black holes\cite{Bekenstein1973,Hawking1975,Solodukhin:2011gn},  given by
 the area of  the event horizon area $A$  in Planck units, implies a discrete holographic  bound on  total information  in any system\cite{tHooft1993,Susskind1995,CohenKaplanNelson1999}. The spatial density of information therefore decreases without bound in large systems. 
To account for the missing information,  space-time itself must have  ``spooky'' nonlocalized  spacelike correlations, like those long familiar in entangled states of quantum particles\cite{RevModPhys.71.S288}.
To avoid information paradoxes, the exotic states in the geometry must entangle, and share their exotic correlations, with the fields that populate it\cite{Hooft:2016cpw}.



Indeed, thermodynamics can be used  to derive all of general relativity---  Einstein's  field equations governing space, time,  gravity and curvature---  as a large scale emergent or collective  behavior of a  quantum system \cite{Jacobson1995,Verlinde2011,Padmanabhan:2013nxa,Jacobson:2015hqa}.
 The underlying states do not resemble particles, waves, pixels, or perturbations of the metric, 
but are delocalized entities that live on null sheets. 
The familiar particle-like graviton modes  of linearized gravity are like phonons in a gas: they do not represent the  fundamental  degrees of freedom, in the same way that quantized sound waves do not represent the fundamental atomic quanta of a gas.
In this picture, even the notion of locality  is emergent.  


 Like general relativity, the thermodynamic theory  by itself does not set a value for the cosmological constant.
 However, in the absence of any form of matter or thermal excitations, it is natural to conjecture that a pure geometrical quantum system in its zero-temperature ground state approaches a  zero-curvature space-time, i.e. $\Lambda\rightarrow 0$ on large scales, albeit with small exotic quantum correlations and their associated fluctuations.

The detailed effect of the exotic geometrical correlations on fields is not known, but  certainly  differs from standard field vacuum fluctuations. In one model\cite{Hogan:2015b,Hogan:2016,Hogan:2017bum}, exotic corrrelations   associated with the  emergence of directions  give rise to new kinds of
exotic rotational  fluctuations in the (also emergent) inertial frame,  which are predicted to identically vanish in standard theory.
Exotic fluctuations in separation $R$ between world lines should  vanish by causal symmetry, but  relative transverse positions 
and directions fluctuate from classical trajectories with variances 
 \begin{equation}\label{transversevariance}
 \langle\Delta x_\perp^2\rangle\equiv k_\perp^{-2}   \approx l_PR \ \ \  {\rm and} \ \ \ 
\langle\Delta\theta^2\rangle\approx l_P/R. 
\end{equation}
Directional  variations on scale $ R$ and timescale $R/c$ produce rotational fluctuations
with a variance in rotation rate
\begin{equation}\label{OmegaR}
\langle\omega^2(R) \rangle \approx    c^2 l_PR^{-3} \approx t_P^{-2} (k_\perp l_P)^6.
\end{equation}
The states live on null cones, so their exotic  spacelike correlations  extend indefinitely, like spooky nonlocal  quantum 
correlations of entangled particle states\cite{RevModPhys.71.S288}. If the new correlations  affect fields everywhere in the same way they do near black holes, they  should   produce detectable signals in new kinds of experiments\cite{holoshear,Holo:Instrument}.

In  emergent quantum gravity with matter fields included, the  behavior of the geometry should be slightly  affected by entanglement with the exotic fluctuations in the field vacuum,
which 
 breaks the exact zero-curvature symmetry of the empty ground state. 
 If the geometry and field systems are maximally entangled, and the total information is shared equally between them,  the spatial information in the field vacuum  has a precisely calculable relationship with the value of the cosmological constant.
 

Suppose  that in the most probable state of the entangled system of fields and geometry, a naturally-flat  geometry adapts to the presence of fields by adopting a tiny emergent curvature--- a cosmological constant--- so that the information is equally shared between matter vacuum and geometry. 
The measured value of $\Lambda$ gives our universe  an event horizon with a radius $c/ H_\Lambda=   1.01\pm 0.024 \times 10^{61} l_P$,
and a total cosmic  information content, exactly analogous to  black hole entropy, 
 \begin{equation}\label{cosmichorizon}
S_\Lambda= \pi t_P^{-2} H_\Lambda^{-2}= 3.2 \pm 0.15 \times 10^{122}.
\end{equation}
The number of degrees of  freedom for a free scalar field with an ultraviolet cutoff at wavenumber $k$ in a 3D volume of radius
$c/ H_\Lambda$   is
${\cal N}_f(k)=(2/ 9\pi) \ k^3 (c/H_\Lambda)^3$.
A free scalar field   therefore matches cosmic information (that is, $S_\Lambda= {\cal N}_f(k)$) for a field cutoff at
\begin{equation}\label{kLambda}
k_\Lambda =  ( H_\Lambda 9\pi^2/2)^{1/3} = 1.65\pm 0.01 \times 10^{-20} l_P^{-1} =  201\pm 1.6\   {\rm MeV}/c\hbar.
\end{equation}

Remarkably, this scale closely matches the field vacuum  we already know about,  the Standard Model.  
It  represents an energy threshold where quantum chromodynamics (QCD) abruptly and spontaneously changes the spatial 
organization and localization of particle states. 
In field theory, the large ratio $\approx 10^{-20}$ is understood to originate from the running coupling constant
of non-Abelian gluon self-interactions\cite{Wilczek1999}.  The interactions blow up at about 200 MeV\cite{Olive2014}, which leads to a phase change in the vacuum, and sets the scale for  the masses,  sizes and interactions of nucleons.  


An important limitation on the precision of this comparison with QCD is the free field approximation used  for the spatial  information content of the field vacuum. Still, even this simplified calculation leads to remarkably good agreement with the measured value of cosmic information--- better than ten percent, as opposed to the disagreement by a factor of $\approx 10^{61}$ found in standard field theory.

In this model, the cosmic acceleration is not actually  constant, but fluctuates in a new way: the entanglement  of emergent quantum geometry with QCD ``shakes the universe apart''.
The radial cosmic acceleration can be visualized as a centrifugal effect of the exotic rotational fluctuations  of the inertial frame, as they ``drag'' the field vacuum.  
The  rotational fluctuations create a kinematical mean square centrifugal acceleration, which can be estimated from Eqs. (\ref{accelerate}), (\ref{OmegaR}),  and (\ref{kLambda}):
\begin{equation}\label{centrifuge}
 \langle H_\Lambda^2\rangle = \langle \ddot r / r \rangle_\Lambda \approx  \langle \omega^2 \rangle_\Lambda \approx t_P^{-2} (k_\Lambda l_P)^6.
\end{equation} 
The coherence scale of the fluctuations is determined by the Chandrasekhar radius for QCD, similar to the size of a neutron star. 
They have a spatial coherence scale  $\approx R \approx k_\Lambda^{-2}\approx 10^{40}$ in Planck units, of the order of 100 km, and
a characteristic frequency $\approx R^{-1}\approx k_\Lambda^2\approx 10^{-40}$ in Planck units,  or about 3000 Hz.

This scenario  suggests that there may be   a precise and quantitative formulation of the long noted coincidence  $\Lambda \approx  m_\pi^6$ (in Planck units),  and also points to some concrete steps  to test the idea.
The effect of exotic  rotational correlations on fields may be directly observable in interferometer cross correlations, using an existing apparatus.
The spatiotemporal fluctuations of acceleration in  the emergent metric  are much harder to observe, but if measured would be a compelling direct cosmic signature.





  \begin{acknowledgments}
This work was supported by the Department
of Energy at Fermilab under Contract No. DE-AC02-07CH11359.  
\end{acknowledgments}

  \bibliographystyle{apsrev4-1.bst}

\bibliography{GRFbib}

\end{document}